\documentclass[preprint]{aastex63}

\shorttitle{HAWC J1825-134}
\graphicspath{{./}{figures/}}
\begin{document}

\title{Evidence of 200 TeV photons from HAWC J1825-134}

\correspondingauthor{Francisco Salesa Greus}
\email{sagreus@ific.uv.es}

\correspondingauthor{Sabrina Casanova}
\email{sabrina.casanova@ifj.edu.pl}

\correspondingauthor{Dezhi Huang}
\email{dezhih@mtu.edu}

\author[0000-0003-0197-5646]{A.~Albert}
\affiliation{Physics Division, Los Alamos National Laboratory, Los Alamos, NM, USA }
\author[0000-0001-8749-1647]{R.~Alfaro}
\affiliation{Instituto de F\'{i}sica, Universidad Nacional Autónoma de México, Ciudad de Mexico, Mexico }
\author{C.~Alvarez}
\affiliation{Universidad Autónoma de Chiapas, Tuxtla Gutiérrez, Chiapas, México}
\author{J.R.~Angeles Camacho}
\affiliation{Instituto de F\'{i}sica, Universidad Nacional Autónoma de México, Ciudad de Mexico, Mexico }
\author{J.C.~Arteaga-Velázquez}
\affiliation{Universidad Michoacana de San Nicolás de Hidalgo, Morelia, Mexico }
\author{K.P.~Arunbabu}
\affiliation{Instituto de Geof\'{i}sica, Universidad Nacional Autónoma de México, Ciudad de Mexico, Mexico }
\author{D.~Avila Rojas}
\affiliation{Instituto de F\'{i}sica, Universidad Nacional Autónoma de México, Ciudad de Mexico, Mexico }
\author[0000-0002-2084-5049]{H.A.~Ayala Solares}
\affiliation{Department of Physics, Pennsylvania State University, University Park, PA, USA }
\author[0000-0003-0477-1614]{V.~Baghmanyan}
\affiliation{Institute of Nuclear Physics Polish Academy of Sciences, PL-31342 IFJ-PAN, Krakow, Poland }
\author[0000-0003-3207-105X]{E.~Belmont-Moreno}
\affiliation{Instituto de F\'{i}sica, Universidad Nacional Autónoma de México, Ciudad de Mexico, Mexico }
\author[0000-0001-5537-4710]{S.Y.~BenZvi}
\affiliation{Department of Physics \& Astronomy, University of Rochester, Rochester, NY , USA }
\author[0000-0002-5493-6344]{C.~Brisbois}
\affiliation{Department of Physics, University of Maryland, College Park, MD, USA }
\author[0000-0003-2158-2292]{T.~Capistrán}
\affiliation{Instituto de Astronom\'{i}a, Universidad Nacional Autónoma de México, Ciudad de Mexico, Mexico }
\author[0000-0002-8553-3302]{A.~Carramiñana}
\affiliation{Instituto Nacional de Astrof\'{i}sica, Óptica y Electrónica, Puebla, Mexico }
\author[0000-0002-6144-9122]{S.~Casanova}
\affiliation{Institute of Nuclear Physics Polish Academy of Sciences, PL-31342 IFJ-PAN, Krakow, Poland }
\author[0000-0002-7607-9582]{U.~Cotti}
\affiliation{Universidad Michoacana de San Nicolás de Hidalgo, Morelia, Mexico }
\author[0000-0002-1132-871X]{J.~Cotzomi}
\affiliation{Facultad de Ciencias F\'{i}sico Matemáticas, Benemérita Universidad Autónoma de Puebla, Puebla, Mexico }
\author[0000-0001-9643-4134]{E.~De la Fuente}
\affiliation{Departamento de F\'{i}sica, Centro Universitario de Ciencias Exactase Ingenierias, Universidad de Guadalajara, Guadalajara, Mexico }
\author{R.~Diaz Hernandez}
\affiliation{Instituto Nacional de Astrof\'{i}sica, Óptica y Electrónica, Puebla, Mexico }
\author[0000-0001-8451-7450]{B.L.~Dingus}
\affiliation{Physics Division, Los Alamos National Laboratory, Los Alamos, NM, USA }
\author[0000-0002-2987-9691]{M.A.~DuVernois}
\affiliation{Department of Physics, University of Wisconsin-Madison, Madison, WI, USA }
\author{M.~Durocher}
\affiliation{Physics Division, Los Alamos National Laboratory, Los Alamos, NM, USA }
\author[0000-0002-0087-0693]{J.C.~Díaz-Vélez}
\affiliation{Departamento de F\'{i}sica, Centro Universitario de Ciencias Exactase Ingenierias, Universidad de Guadalajara, Guadalajara, Mexico }
\author[0000-0001-5737-1820]{K.~Engel}
\affiliation{Department of Physics, University of Maryland, College Park, MD, USA }
\author[0000-0001-7074-1726]{C.~Espinoza}
\affiliation{Instituto de F\'{i}sica, Universidad Nacional Autónoma de México, Ciudad de Mexico, Mexico }
\author{K.~Fang}
\affiliation{Department of Physics, Stanford University: Stanford, CA 94305–4060, USA}
\author[0000-0002-0794-8780]{H.~Fleischhack}
\affiliation{Department of Physics, Michigan Technological University, Houghton, MI, USA }
\author{N.~Fraija}
\affiliation{Instituto de Astronom\'{i}a, Universidad Nacional Autónoma de México, Ciudad de Mexico, Mexico }
\author{A.~Galván-Gámez}
\affiliation{Instituto de Astronom\'{i}a, Universidad Nacional Autónoma de México, Ciudad de Mexico, Mexico }
\author{D.~Garcia}
\affiliation{Instituto de F\'{i}sica, Universidad Nacional Autónoma de México, Ciudad de Mexico, Mexico }
\author[0000-0002-4188-5584]{J.A.~García-González}
\affiliation{Instituto de F\'{i}sica, Universidad Nacional Autónoma de México, Ciudad de Mexico, Mexico }
\author[0000-0003-1122-4168]{F.~Garfias}
\affiliation{Instituto de Astronom\'{i}a, Universidad Nacional Autónoma de México, Ciudad de Mexico, Mexico }
\author{G.~Giacinti}
\affiliation{Max-Planck Institute for Nuclear Physics, 69117 Heidelberg, Germany}
\author[0000-0002-5209-5641]{M.M.~González}
\affiliation{Instituto de Astronom\'{i}a, Universidad Nacional Autónoma de México, Ciudad de Mexico, Mexico }
\author[0000-0002-9790-1299]{J.A.~Goodman}
\affiliation{Department of Physics, University of Maryland, College Park, MD, USA }
\author{J.P.~Harding}
\affiliation{Physics Division, Los Alamos National Laboratory, Los Alamos, NM, USA }
\author{B.~Hona}
\affiliation{Department of Physics and Astronomy, University of Utah, Salt Lake City, UT, USA }
\author[0000-0002-5447-1786]{D.~Huang}
\affiliation{Department of Physics, Michigan Technological University, Houghton, MI, USA }
\author[0000-0002-5527-7141]{F.~Hueyotl-Zahuantitla}
\affiliation{Universidad Autónoma de Chiapas, Tuxtla Gutiérrez, Chiapas, México}
\author{P.~Hüntemeyer}
\affiliation{Department of Physics, Michigan Technological University, Houghton, MI, USA }
\author[0000-0001-5811-5167]{A.~Iriarte}
\affiliation{Instituto de Astronom\'{i}a, Universidad Nacional Autónoma de México, Ciudad de Mexico, Mexico }
\author[0000-0002-6738-9351]{A.~Jardin-Blicq}
\affiliation{Max-Planck Institute for Nuclear Physics, 69117 Heidelberg, Germany}
\affiliation{Department of Physics, Faculty of Science, Chulalongkorn University, 254
Phayathai Road,Pathumwan, Bangkok 10330, Thailand}
\affiliation{National Astronomical Research Institute of Thailand (Public
Organization), Don Kaeo, MaeRim, Chiang Mai 50180, Thailand}
\author[0000-0003-4467-3621]{V.~Joshi}
\affiliation{Erlangen Centre for Astroparticle Physics, Friedrich-Alexander-Universit\"at Erlangen-N\"urnberg, Erlangen, Germany}
\author{G.J.~Kunde}
\affiliation{Physics Division, Los Alamos National Laboratory, Los Alamos, NM, USA }
\author{A.~Lara}
\affiliation{Instituto de Geof\'{i}sica, Universidad Nacional Autónoma de México, Ciudad de Mexico, Mexico }
\author[0000-0002-2467-5673]{W.H.~Lee}
\affiliation{Instituto de Astronom\'{i}a, Universidad Nacional Autónoma de México, Ciudad de Mexico, Mexico }
\author[0000-0001-5516-4975]{H.~León Vargas}
\affiliation{Instituto de F\'{i}sica, Universidad Nacional Autónoma de México, Ciudad de Mexico, Mexico }
\author{J.T.~Linnemann}
\affiliation{Department of Physics and Astronomy, Michigan State University, East Lansing, MI, USA }
\author[0000-0001-8825-3624]{A.L.~Longinotti}
\affiliation{Instituto de Astronom\'{i}a, Universidad Nacional Autónoma de México, Ciudad de Mexico, Mexico }
\affiliation{Instituto Nacional de Astrof\'{i}sica, Óptica y Electrónica, Puebla, Mexico }
\author[0000-0003-2810-4867]{G.~Luis-Raya}
\affiliation{Universidad Politecnica de Pachuca, Pachuca, Hgo, Mexico }
\author{J.~Lundeen}
\affiliation{Department of Physics and Astronomy, Michigan State University, East Lansing, MI, USA }
\author[0000-0001-8088-400X]{K.~Malone}
\affiliation{Physics Division, Los Alamos National Laboratory, Los Alamos, NM, USA }
\author[0000-0001-9077-4058]{V.~Marandon}
\affiliation{Max-Planck Institute for Nuclear Physics, 69117 Heidelberg, Germany}
\author[0000-0001-9052-856X]{O.~Martinez}
\affiliation{Facultad de Ciencias F\'{i}sico Matemáticas, Benemérita Universidad Autónoma de Puebla, Puebla, Mexico }
\author{J.~Martínez-Castro}
\affiliation{Centro de Investigaci\'on en Computaci\'on, Instituto Polit\'ecnico Nacional, M\'exico City, M\'exico.}
\author[0000-0002-2610-863X]{J.A.~Matthews}
\affiliation{Dept of Physics and Astronomy, University of New Mexico, Albuquerque, NM, USA }
\author[0000-0002-8390-9011]{P.~Miranda-Romagnoli}
\affiliation{Universidad Autónoma del Estado de Hidalgo, Pachuca, Mexico }
\author[0000-0002-1114-2640]{E.~Moreno}
\affiliation{Facultad de Ciencias F\'{i}sico Matemáticas, Benemérita Universidad Autónoma de Puebla, Puebla, Mexico }
\author[0000-0002-7675-4656]{M.~Mostafá}
\affiliation{Department of Physics, Pennsylvania State University, University Park, PA, USA }
\author[0000-0003-0587-4324]{A.~Nayerhoda}
\affiliation{Institute of Nuclear Physics Polish Academy of Sciences, PL-31342 IFJ-PAN, Krakow, Poland }
\author[0000-0003-1059-8731]{L.~Nellen}
\affiliation{Instituto de Ciencias Nucleares, Universidad Nacional Autónoma de Mexico, Ciudad de Mexico, Mexico }
\author[0000-0001-9428-7572]{M.~Newbold}
\affiliation{Department of Physics and Astronomy, University of Utah, Salt Lake City, UT, USA }
\author[0000-0002-6859-3944]{M.U.~Nisa}
\affiliation{Department of Physics and Astronomy, Michigan State University, East Lansing, MI, USA }
\author[0000-0001-7099-108X]{R.~Noriega-Papaqui}
\affiliation{Universidad Autónoma del Estado de Hidalgo, Pachuca, Mexico }
\author[0000-0002-5448-7577]{N.~Omodei}
\affiliation{Department of Physics, Stanford University: Stanford, CA 94305–4060, USA}
\author{A.~Peisker}
\affiliation{Department of Physics and Astronomy, Michigan State University, East Lansing, MI, USA }
\author[0000-0002-8774-8147]{Y.~Pérez Araujo}
\affiliation{Instituto de Astronom\'{i}a, Universidad Nacional Autónoma de México, Ciudad de Mexico, Mexico }
\author[0000-0001-5998-4938]{E.G.~Pérez-Pérez}
\affiliation{Universidad Politecnica de Pachuca, Pachuca, Hgo, Mexico }
\author[0000-0002-6524-9769]{C.D.~Rho}
\affiliation{Natural Science Research Institute, University of Seoul, Seoul, Republic of Korea}
\author[0000-0003-1327-0838]{D.~Rosa-González}
\affiliation{Instituto Nacional de Astrof\'{i}sica, Óptica y Electrónica, Puebla, Mexico }
\author{H.~Salazar}
\affiliation{Facultad de Ciencias F\'{i}sico Matemáticas, Benemérita Universidad Autónoma de Puebla, Puebla, Mexico }
\author[0000-0002-8610-8703]{F.~Salesa Greus}
\affiliation{Institute of Nuclear Physics Polish Academy of Sciences, PL-31342 IFJ-PAN, Krakow, Poland }
\affiliation{Instituto de Física Corpuscular, CSIC, Universitat de València, E-46980, Paterna, Valencia, Spain}
\author{A.~Sandoval}
\affiliation{Instituto de F\'{i}sica, Universidad Nacional Autónoma de México, Ciudad de Mexico, Mexico }
\author{M.~Schneider}
\affiliation{Department of Physics, University of Maryland, College Park, MD, USA }
\author{F.~Serna}
\affiliation{Instituto de F\'{i}sica, Universidad Nacional Autónoma de México, Ciudad de Mexico, Mexico }
\author[0000-0002-1492-0380]{R.W.~Springer}
\affiliation{Department of Physics and Astronomy, University of Utah, Salt Lake City, UT, USA }
\author[0000-0001-9725-1479]{K.~Tollefson}
\affiliation{Department of Physics and Astronomy, Michigan State University, East Lansing, MI, USA }
\author[0000-0002-1689-3945]{I.~Torres}
\affiliation{Instituto Nacional de Astrof\'{i}sica, Óptica y Electrónica, Puebla, Mexico }
\author{R.~Torres-Escobedo}
\affiliation{Departamento de F\'{i}sica, Centro Universitario de Ciencias Exactase Ingenierias, Universidad de Guadalajara, Guadalajara, Mexico }
\affiliation{Department of Physics and Astronomy, Texas Tech University, USA}
\author{F.~Ureña-Mena}
\affiliation{Instituto Nacional de Astrof\'{i}sica, Óptica y Electrónica, Puebla, Mexico }
\author[0000-0001-6876-2800]{L.~Villaseñor}
\affiliation{Facultad de Ciencias F\'{i}sico Matemáticas, Benemérita Universidad Autónoma de Puebla, Puebla, Mexico }
\author{E.~Willox}
\affiliation{Department of Physics, University of Maryland, College Park, MD, USA }
\author{H.~Zhou}
\affiliation{Tsung-Dao Lee Institute \& School of Physics and Astronomy, Shanghai Jiao Tong University, Shanghai, China}
\author{C.~de León}
\affiliation{Universidad Michoacana de San Nicolás de Hidalgo, Morelia, Mexico }



\begin{abstract}

The Earth is bombarded by ultra-relativistic particles, known as cosmic rays (CRs). CRs with energies up to a few PeV (=10$^{15}$ eV), the knee in the particle spectrum, are believed to have a Galactic origin. One or more factories of PeV CRs, or PeVatrons, must thus be active within our Galaxy. The direct detection of PeV protons from their sources is not possible since they are deflected in the Galactic magnetic fields. Hundred TeV $\gamma$-rays from decaying $\pi^0$, produced when PeV CRs collide with the ambient gas, can provide the decisive evidence of proton acceleration up to the knee. Here we report the discovery by the High Altitude Water Cherenkov (HAWC) observatory of the $\gamma$-ray source, HAWC~J1825-134, whose energy spectrum extends well beyond 200 TeV without a break or cutoff. The source is found to be coincident with a giant molecular cloud. The ambient gas density is as high as 700 protons/cm$^3$. While the nature of this extreme accelerator remains unclear, CRs accelerated to energies of several PeV colliding with the ambient gas likely produce the observed radiation.

\end{abstract}



\section{Introduction} 
\label{sec:intro}

A very prominent feature in the CR spectrum measured at Earth is the spectral steepening at several PeV known as the knee, which is believed to mark the maximum acceleration energy achievable in the Galaxy. A single population of CR accelerators, namely young supernova remnants (SNRs), has long been thought to support the entire Galactic CR population up to the knee. An efficient mechanism which can explain sub-PeV to PeV particle acceleration has not been found for historical and older SNR shells, but it might be found for much younger SNe. From an observational point of view the $\gamma$-ray spectra of young SNRs, with the possible exception of the SNR G106.3+2.7 \citep{Albert:2020ngw}, show clear cutoffs in the TeV range \citep{Aharonian:2018oau}. SNRs are proven to be efficient CR accelerators only up to about 100 TeV.  Observational evidence of cosmic rays acceleration up to PeV energies has been found within the Galactic Centre (GC) region. The GC source, whose nature is still debated, can account only for a small portion of the CR energy budget in the Galaxy~\citep{Abramowski:2016mir}. This leaves space for other accelerators or classes of accelerators such as star forming regions or associations of stars to be major contributors to the Galactic CR population~\citep{Cesarski,Parizot:2004em,Aharonian:2018oau}. 

HAWC%\footnote{\url{https://www.hawc-observatory.org/}} 
~is a cosmic and $\gamma$-ray observatory located at a latitude of 19$^{\circ}$ North and an altitude of 4100~m on the slope of the Sierra Negra volcano close to Puebla, in central Mexico~\citep{Abeysekara:2017mjj}.
The HAWC detector has a large field of view (2 sr), high duty cycle (\textgreater 90\% uptime), an angular resolution which can be as good as 0.1$^{\circ}$~\citep{Abeysekara:2019edl}, and unprecedented sensitivity for $\gamma$-ray sources beyond 100 TeV. HAWC has carried out the deepest ever survey of the Galactic Plane in the
unexplored range of $\gamma$-rays with energies beyond 50 TeV, which is crucial for the search of Galactic PeVatrons~\citep{Abeysekara:2019gov}. Among the HAWC high energy sources, eHWC J1825-134, stands out for having the highest flux above 10 TeV. The new analysis of the eHWC J1825-134 region, presented in Section~\ref{sec:analysis} and \ref{sec:results}, extends the previous analysis by investigating the morphology and spectrum of this region in more detail. A discussion on the possible emission mechanisms and the possible counterparts of the HAWC radiation is presented in Section \ref{sec:discussion}. Finally, conclusions are given in Section~\ref{sec:conclusions}.

%%%%%%%%%%%%%%%%%%%%%%%%%%%%

\begin{figure}[ht!]
\begin{center}
\resizebox{1.0\textwidth}{!}{%
\includegraphics[width=0.5\textwidth]{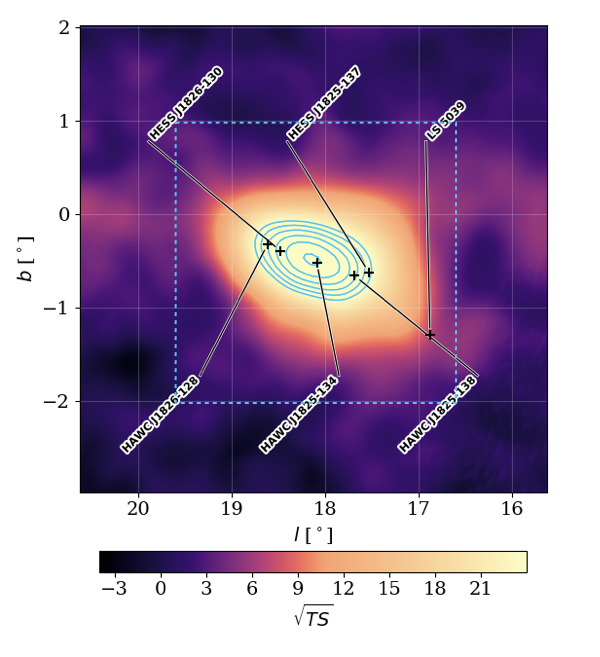}%
\quad
\includegraphics[width=0.5\textwidth]{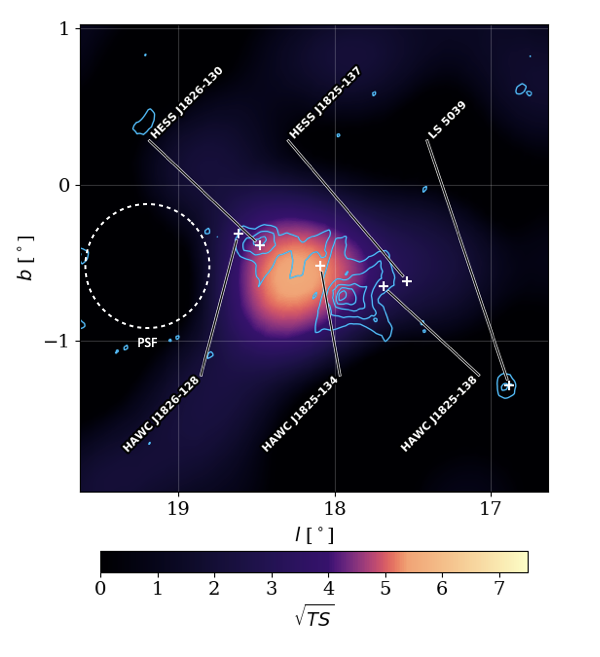}%
}
\caption{Left: significance map around the eHWC J1825-134 region for reconstructed energies greater than 1 TeV. TS refers to the likelihood ratio Test Statistic described in Equation~\ref{ts} in Section~\ref{sec:results}. The blue significance contours correspond to $\rm{\sqrt{TS}}$ at 26, 28 30, 32 and 34. Right: zoom in of the blue dashed region (from left map) including only reconstructed energies greater than 177 TeV. The blue contours corresponding to H.E.S.S. excess at 20, 35, 50, 60 and 70 counts with energies beyond 10 TeV~\citep{Abdalla:2018qgt}. The H.E.S.S. sources location are taken from~\citep{H.E.S.S.:2018zkf}. The white dashed circle represents the 68\% containment of the point spread function (PSF) above 177~TeV obtained from simulation. Significance maps are made assuming a point-source hypothesis and a power-law spectrum with -2.6 index. HAWC does not detect significant emission from LS 5039.}
    
\label{fig: map}
\end{center}
\end{figure}

%%%%%%%%%%%%%%%%%%%%%%%%%%%%

%%%%%%%%%%%%%%%%%%%%%%%%%%%
\begin{figure}[htpb]
\begin{center}
\resizebox{1.0\textwidth}{!}{%
\includegraphics[width=0.7\textwidth]{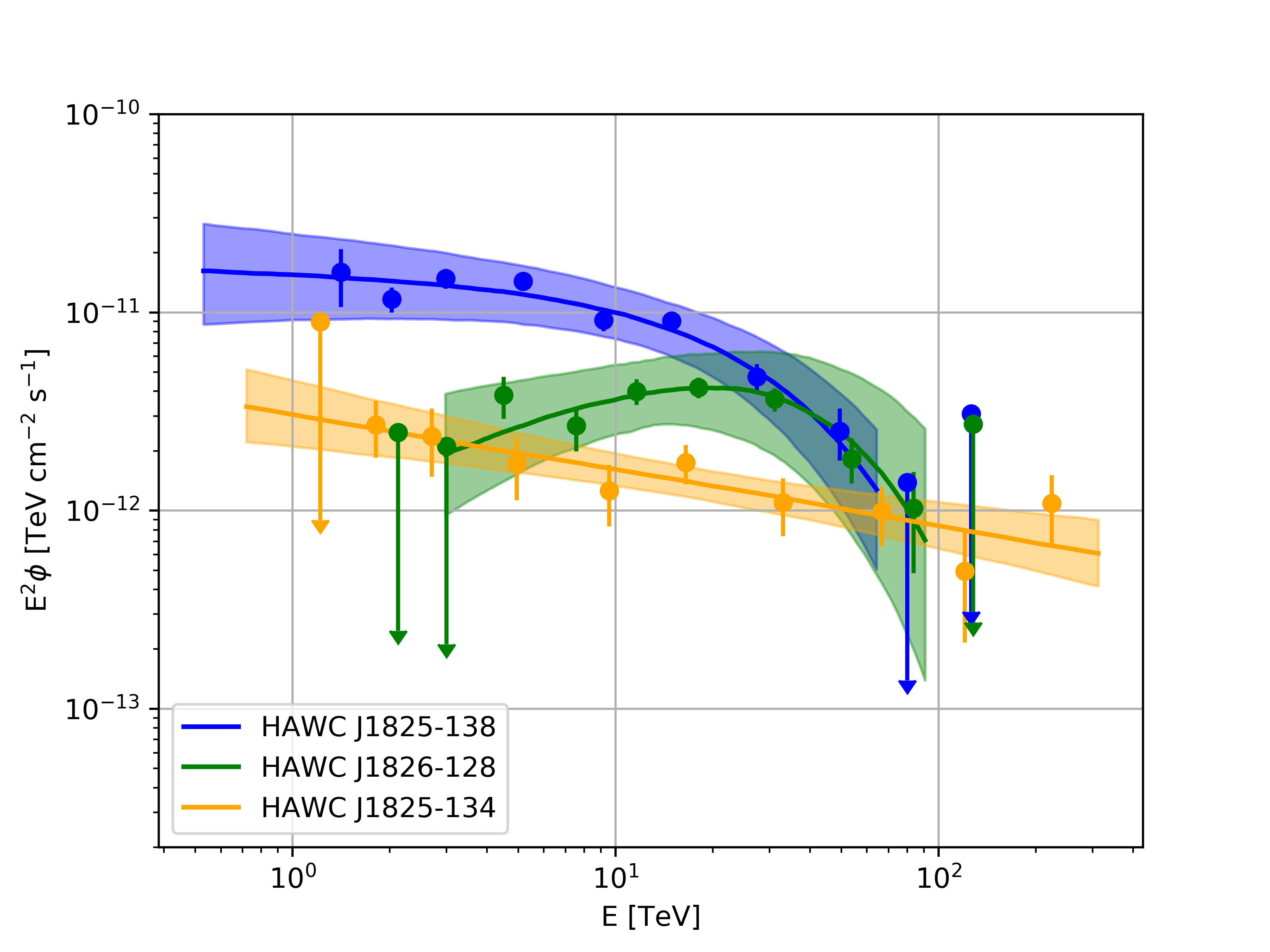}%
}
\caption{$\gamma$-ray spectral energy distribution for the sources in the region of interest around eHWC~J1825-134 that were considered in the model. For more details see Section~\ref{sec:results}.}
    
\label{fig: spectra}
\end{center}
\end{figure}   
%%%%%%%%%%%%%%%%%%%%%%%%%%%%
\section{Analysis} \label{sec:analysis}

Recently, the HAWC energy resolution has improved thanks to the implementation of two independent energy-estimation methods, which makes it possible to measure $\gamma$-ray energies beyond 100 TeV~\citep{Abeysekara:2019edl}.
Based on this improvement the collaboration released a high-energy catalogue of sources emitting above 56~TeV~\citep{Abeysekara:2019gov}.
In this catalogue, three sources showed significant emission above 100 TeV. One of these sources is eHWC J1825-134 which is located within a complex region containing more than one source. The analysis presented in this work resolves and characterizes the different sources contained in the region, thanks to the HAWC angular resolution, which is best at high energies.

We performed a Maximum Likelihood analysis using the Multi-Mission Maximum Likelihood (3ML)~\citep{Vianello:2015wwa} software framework.
The dataset used for the analysis contains 1343 days of observations. This means 300 days more than the dataset used for the high-energy catalogue~\citep{Abeysekara:2019gov}, and almost three times the live time of the second HAWC catalogue~\citep{Abeysekara:2017hyn} dataset where the source was identified as 2HWC J1825-134. In order to assess the significance of the detected sources we used the likelihood ratio test statistic (TS) defined as:
\begin{equation}\label{ts}
    TS=2~\ln\bigg(\frac{L_{alt}}{L_{null}}\bigg)
\end{equation}
where, $L_{alt}$ is the maximum likelihood of the alternative hypothesis~(background + source model) and $L_{null}$ is the likelihood of the null hypothesis~(background only).

We make use of the ground parameter energy estimation method introduced in~\citep{Abeysekara:2019edl}. The region of interest (ROI) used in the analysis is a 2.5$^{\circ}$ disk around the position of eHWC J1825-134. In order to reproduce the observed emission several models were tested. The one showing the best agreement in terms of statistical significance has the following components. First, two extended Gaussian-shaped $\gamma$-ray sources, with a power law with an exponential cut-off as $\gamma$-ray spectrum
\begin{equation}\label{spt}
\phi(E) = \phi_{0} \bigg(\frac{E}{E_{0}}\bigg)^{-\alpha} \exp\bigg(\frac{-E}{E_{\rm cut}}\bigg),
\end{equation}

%$\gamma$-ray spectrum, $\phi(E) = \phi_{0} \bigg(\frac{E}{E_{0}}\bigg)^{-\alpha} \exp\bigg(\frac{-E}{E_{\rm cut}}\bigg)$, 
\noindent which are compatible in position with H.E.S.S. sources (H.E.S.S. J1826-130 and H.E.S.S. J1825-137). Secondly, a point source with a simple power-law spectrum. The nearby TeV $\gamma$-ray binary LS 5039 %at [RA= 276.56$^{\circ}$, Dec= -14.83$^{\circ}$], which happens to be within the ROI, 
was not included in the model because its average flux is below the HAWC sensitivity and it was not significantly detected. Finally, the model also includes a Galactic diffuse emission (GDE) component, with a Gaussian shape along the Galactic latitude (b) with a fixed width of 1$^{\circ}$ and peaked at b = 0$^{\circ}$ based on previous HAWC studies~\citep{Rho:2017yky}. 

\section{Results}\label{sec:results}

 The HAWC significance map for the 1 TeV to 316 TeV energy range is plotted in Figure~\ref{fig: map} left. The HAWC significance map of the region for events above 177 TeV only is shown in Figure~\ref{fig: map} right. The $\gamma$-ray emission in the 1 TeV to 350 TeV energy range observed by HAWC from eHWC~J1825-134 is well described by four components including galactic diffuse emission (GDE), the likely counterparts to known $\gamma$-ray sources, HESS J1825-137 and HESS J1826-130, and a newly discovered source, HAWC J1825-134. As shown in Figure~\ref{fig: spectra}, HAWC~J1825-134, presents a hard power-law energy spectrum with index -2.28$\pm$0.12~(stat.)$^{+0.11}_{-0.04}$ (syst.). The spectrum continues beyond 200 TeV without evidence of a cutoff or break.

To confirm the detection of the point source HAWC J1825-134, a likelihood ratio test is performed between the model using 3 sources, introduced in the previous section, and a 2-source model, similar to the one presented in~\citep{SalesaGreus:2020mpf}, which does not include HAWC J1825-134. According to Wilk's theorem ~\citep{Wilks:1938dza}, for nested models by assuming the 2-source model is true, the TS is distributed as $\chi^{2}$distribution. We define the test statistic difference:
\begin{equation}\label{dts}
    \Delta TS=2~\ln\bigg(\frac{L_{3-source}}{L_{2-source}}\bigg)
\end{equation}
where $L_{3-source}$ is the maximum likelihood of the 3-source model and $L_{2-source}$ is the maximum likelihood of the 2-source model. The test result shows the 3-source model is favored by $\Delta$TS~=~38. Additionally, we perform a Monte Carlo study, from which we estimate that the $\Delta$TS of 38 corresponds to 5.2 sigma after accounting for trials (Section~\ref{sec:modeling}).

The results of the parameters from the maximum likelihood analysis are summarized in Table~\ref{tab: table}. The TS value in the table refers only to the improvement added by that particular source, i.e., it considers all the other sources (including the GDE) as the null hypothesis.
For the sources modeled in the analysis we use a pivot energy, E$_{0}$ in Equation~\ref{spt}, of 18~TeV.

%%%%%%%%%  TABLE  %%%%%%%%%%%%
\begin{table}[htpb]
\scriptsize
%\footnotesize
%\small
\renewcommand{\arraystretch}{2}
  \begin{center}
    \begin{tabular} { |c|c|c|c|c|c|c|c| } 
      \hline
      Source Name & RA [$^{\circ}$] & dec [$^{\circ}$] & width [$^{\circ}$] & $\phi_{0}$ [$\rm cm^{-2} TeV^{-1} s^{-1}$] & $\alpha$ & E$_{\rm cut}$ [TeV] & TS \\
      \hline
      HAWC J1825-138 & 276.38 $^{+0.04}_{-0.04}$ & -13.86 $^{+0.05}_{-0.05}$ & 0.47 $^{+0.04 ~ +0.02}_{-0.04 ~ -0.05}$ & 4.5$^{+1.4 ~ +1.1}_{-1.0 ~ -2.0} \times 10^{-14}$ & 2.02 $^{+0.15 ~ +0.19}_{-0.15 ~ -0.27}$ & 27$^{+9 ~ +12}_{-7 ~ -4}$ & 142\\
      \hline
      HAWC J1826-128 & 276.50 $^{+0.03}_{-0.03}$ & -12.86 $^{+0.04}_{-0.04}$ & 0.20 $^{+0.03 ~ +0.00}_{-0.03 ~ -0.02}$ & 2.7$^{+1.1 ~ +1.3}_{-0.8 ~ -1.4} \times 10^{-14}$ & 1.2 $^{+0.4 ~ +0.4}_{-0.4 ~ -0.5}$ & 24$^{+10 ~ +15}_{-7 ~ -7}$& 83\\ 
      \hline
      HAWC J1825-134 & 276.44$^{+0.03}_{-0.03}$ & -13.42 $^{+0.04}_{-0.04}$ & -- & 4.2$^{+0.8 ~ +1.1}_{-0.7 ~ -1.5} \times 10^{-15}$ & 2.28 $^{+0.12 ~ +0.10}_{-0.12 ~ -0.04}$ & -- & 38\\ 
      \hline
      
    \end{tabular}
  \end{center}
  \caption{Values of the spectral parameters from the likelihood fit for each of the sources in the model: RA (right ascension), dec (declination), Gaussian width for the extended sources, $\gamma$-ray flux at the pivot energy, spectral index, energy cutoff and TS. The uncertainties both statistical and systematic, in that order, are quoted.}
  \label{tab: table}
\end{table}
%%%%%%%%%%%%%%%%%%%%%%%%%%%%%%

The differential flux at 18 TeV of the new source, HAWC J1825-134, is 4.2$^{+0.8}_{-0.7}$
(stat.)~$^{+1.0}_{-1.5}$ (syst.) $\times$ 10$^{-15}$ cm$^{-2}$ s$^{-1}$ TeV$^{-1}$ and its spectrum is a pure power law with spectral index 2.28 $\pm$ 0.12 (stat.)~$^{+0.10}_{-0.04}$ (syst.). Its integral flux above 1 TeV is 2.4 $\times$ 10$^{-12}$ cm$^{-2}$ s$^{-1}$, corresponding to roughly 12 $\%$ of the Crab flux above 1 TeV. The source is considered as a point source with an upper limit of its Gaussian width extension of 0.18$^{\circ}$ at 95\% CL.

The spectral shape of the GDE is assumed to follow a simple power law.
It is important to notice that the GDE component of the model adopted includes not only the emission from the interactions between CRs and interstellar gas/radiation fields, but also all the emission from sources below the HAWC detection threshold. For a pivot energy of E$_{0}$ = 7 TeV, the best values for the fitted GDE parameters are: spectral index $\alpha = 2.61\pm 0.06$ (stat.) $^{+0.04}_{-0.02}$ (syst.), and flux normalization $\phi_{0} = 5.2\pm 0.6$ (stat.)$^{+1.5}_{-0.7}$ (syst.)$\times 10^{-11} ~ \rm cm^{-2} TeV^{-1}s^{-1}sr^{-1}$. The TS for the GDE component is 62.

The result of the likelihood fit is susceptible to the effects of systematic errors. Based on previous HAWC studies we considered the following effects: angular resolution mismodeling, late light simulation, charge uncertainty, absolute PMT efficiency, PMT threshold.
For further details on each of these effects see~\citep{Abeysekara:2017mjj,Abeysekara:2019edl}. We changed the instrument response, which is modeled using Monte Carlo simulations, to probe the effects mentioned before. Apart from those effects, we also considered the impact of an asymmetric PSF on the total flux measured using simulations. To get a total systematic uncertainty, all possible systematic uncertainties were added in quadrature. Finally, studies done for the third HAWC catalogue~\citep{Albert:2020fua} showed that there is a systematic error on the absolute pointing of the order of 0.2$^{\circ}$ for the region studied here. This systematic is expected to shift the location of the sources in our analysis in a single direction. In any case, even a shift of 0.5 deg would not change the conclusion of the paper, as it is discussed later. We also tested the systematic uncertainties coming from a possible mismodeling of the morphology of HAWC~J1825-138, and in particular the impact on HAWC J1825-134. The study indicates that this effect is smaller than other detector systematic uncertainties. 
%For the morphology systematic study, the asymmetric 
%model for HAWC J1825-138 was tested. The result indicates the asymmetric model is not preferred in HAWC data. Also, the shifts in flux are less than detector systematic uncertain.

All the spectra of the relevant sources are plotted in Figure~\ref{fig: spectra}. The flux points are obtained following the prescription presented in~\citep{Abeysekara:2019edl} where, for a given energy range, the flux normalization is fitted while keeping fixed the other parameters in the model. 
When the TS of a given point is below 4, then an upper limit is set for this point at 90\% CL using the Profile Likelihood method~\citep{Venzon:1988}.
The energy ranges, for which the uncertainty bands are displayed, are computed using the hard-cutoff method explained in~\citep{Abeysekara:2017mjj} for a 68\% CL.
Using the same method we can test the maximum energy of the new source, HAWC J1825-134, for which the likelihood fit shows no evidence of a spectral cutoff or break. In this case, the lower limits on the maximum photon energy are 163$^{+18}_{-22}$ (syst.), 209$^{+39}_{-12}$ (syst.), and 312$^{+20}_{-24}$ (syst.) TeV for CL of 99.7, 95, and 68\% respectively. 
These values are consistent with the ones reported in another HAWC analysis~\citep{Albert:2019nnn} using a different energy estimator and modeling.

\section{Discussion} \label{sec:discussion}

The TeV $\gamma$-ray emission from HAWC J1825-134 can in principle be produced by either PeV CR protons colliding with the ambient gas (hadronic mechanism) or by sub-PeV to PeV electrons upscattering the cosmic microwave background (CMB) photons through Inverse Compton scattering (IC) (leptonic mechanism). 

There are however several arguments why a major contribution to the emission of HAWC~J1825-134 from the leptonic mechanism is unlikely.  To reach energies of 200 TeV in up-scattered photons as measured by HAWC, we have to assume electron energies of at least 200 TeV or more, the median energy of these electrons being roughly 500 TeV \citep{Aharonian:2004yt}. The leptonic population has in fact no cutoff (Section~\ref{sec:location}). %Section~\ref{sec:results}{Appendix}).
These electrons could be in principle accelerated either in nearby pulsars or SNRs. 

In SNRs the acceleration time of 500 TeV electrons in the Bohm regime and for fast shock speed, $\rm{v_{sh}}$ of 2000 km/s, would be $ \rm {t_{acc}} \simeq   \rm{ \frac{D(E)}{{v_{sh}}^2}}$ , where  $D(E) = \rm{ \frac{ \eta c R_{larmor}} {3}}$ and $\eta$=1 for Bohm diffusion. The acceleration time, ${t_{acc}} \simeq  1.5 \times 10^5  \rm {(\frac{E}{500TeV})} \rm{{(\frac{B}{1 \mu G})}^{-1}} \rm{{(\frac{v_{sh}}{2000 km/s})}^{-2}}$ yr \citep{Aharonian:2004yt}, would thus be longer than the ages of any nearby SNR. The way to shorten the acceleration time would be to increase the B field by at least a factor of 100-1000, then $ \rm {t_{acc}} \sim$ 10$^3$-10$^2$ yr. However, in this case the synchrotron cooling time, $ \rm{t_{cool} \simeq 2 \, {(\frac{B} {100 \mu G})}^{-2} {(\frac{E} {500 TeV})}^{-1}}$ yr would be extremely short, of the order of two years or even less. Effective acceleration implies strongly amplified magnetic fields, which in turn cause the electron synchrotron cooling time to be extremely short. Thus, SNR acceleration of electrons is disfavored by acceleration and cooling time arguments.

If nearby pulsars accelerated electrons up to 500 TeV, the shortest acceleration time would be $ \rm {t_{acc} = \frac{R_{larmor}} {c} \simeq  2 {(\frac{E}{500 TeV})} {(\frac{B}{1 \mu G})}^{-1} }$ yr for extreme accelerators working at the highest possible rate in Bohm regime \citep{Abramowski:2016mir}. These electrons would have to diffuse from the pulsars to the entire HAWC J1825-134 region. The magnetic field B in dense gas regions such as HAWC J1825-134 is known from Zeeman measurements in molecular cloud cores \citep{Crutcher}. The  correlation between magnetic field strength and gas density is B$\sim 100(\frac{n}{10^4 cm^{-3}})^{0.5}$ $\mu$G. Within HAWC J1825-134 the average gas density is n=700 cm$^{-3}$ (see Section~\ref{sec:gasdensity}), which corresponds to an average B $\simeq$ 27 $\mu$G. The cooling time and the distance travelled by the electrons are both proportional to $(\frac{n}{10^4 cm^{-3}})^{-1}$. The cooling time for 500 TeV electrons in a B field of 27 $\mu$G would be roughly 28 yr, giving a maximum range of about 8 pc for particles travelling at the light speed, whereas the distance to travel from the closest pulsar, PSR J1826-1334, is of 30 pc.  Furthermore, the regime of particle transport depends upon the level of turbulence. In turbulent HII regions, such as the region of HAWC J1825-134, one expects the particle diffusion coefficient to be strongly suppressed, so that particles diffuse at least tens to hundreds of times slower than in the average Galactic interstellar medium. In summary, the escape of such energetic electrons from the accelerator and their propagation from the local pulsars to the region of HAWC J1825-134 would require weak magnetic fields and extremely fast diffusion, which are unrealistic conditions in regions of high gas density (for further information see Section~\ref{sec:gasdensity}). In Section~\ref{sec:location} we discuss the effect of the uncertainty on the HAWC J1825-134 fit location for the leptonic scenario.

The major contribution to the emission from HAWC J1825-134 likely comes from PeV protons colliding off the ambient gas and producing $\pi^0$ which immediately decay into hundred TeV photons. The spectrum of the CR population producing the HAWC emission is a pure power law with spectral index -2.30 $\pm$ 0.12 and without cutoff, as obtained with a Naima fit~\citep{Zabalza:2015bsa}. A lower limit for the maximum energy of the protons producing the HAWC J1825-134 emission is 1.3 PeV at 95\% confidence level (CL). The total CR energy budget required to explain the HAWC radiation above 10 TeV is W$_p$ = 1.2 $\times$ 10$^{47} {(\frac{\rm n}{700 {\rm cm^{-3}}})}^{-1}$ erg. For an estimated ambient gas density of 700 cm$^{-3}$ (see  Section~\ref{sec:gasdensity}) the cosmic ray energy density above 10 TeV, w$_p$, is 0.3 eV/cm$^3$, which is about 300 times higher than the locally measured CR energy density above 10 TeV \citep{Adriani:2019aft}. Such a high cosmic ray density is likely produced by a currently 
operating accelerator inside HAWC J1825-134 or very close to the HAWC source. Extrapolating the proton spectrum to GeV energies, the required energy budget would be W$_p$ = 3.4 $\times$ 10$^{48} {(\frac{\rm n}{700 {\rm cm^{-3}}})}^{-1}$ erg above 1 GeV.

\begin{figure}[htpb]
\begin{center}
\resizebox{1.0\textwidth}{!}{%
\includegraphics[width=0.7\textwidth]{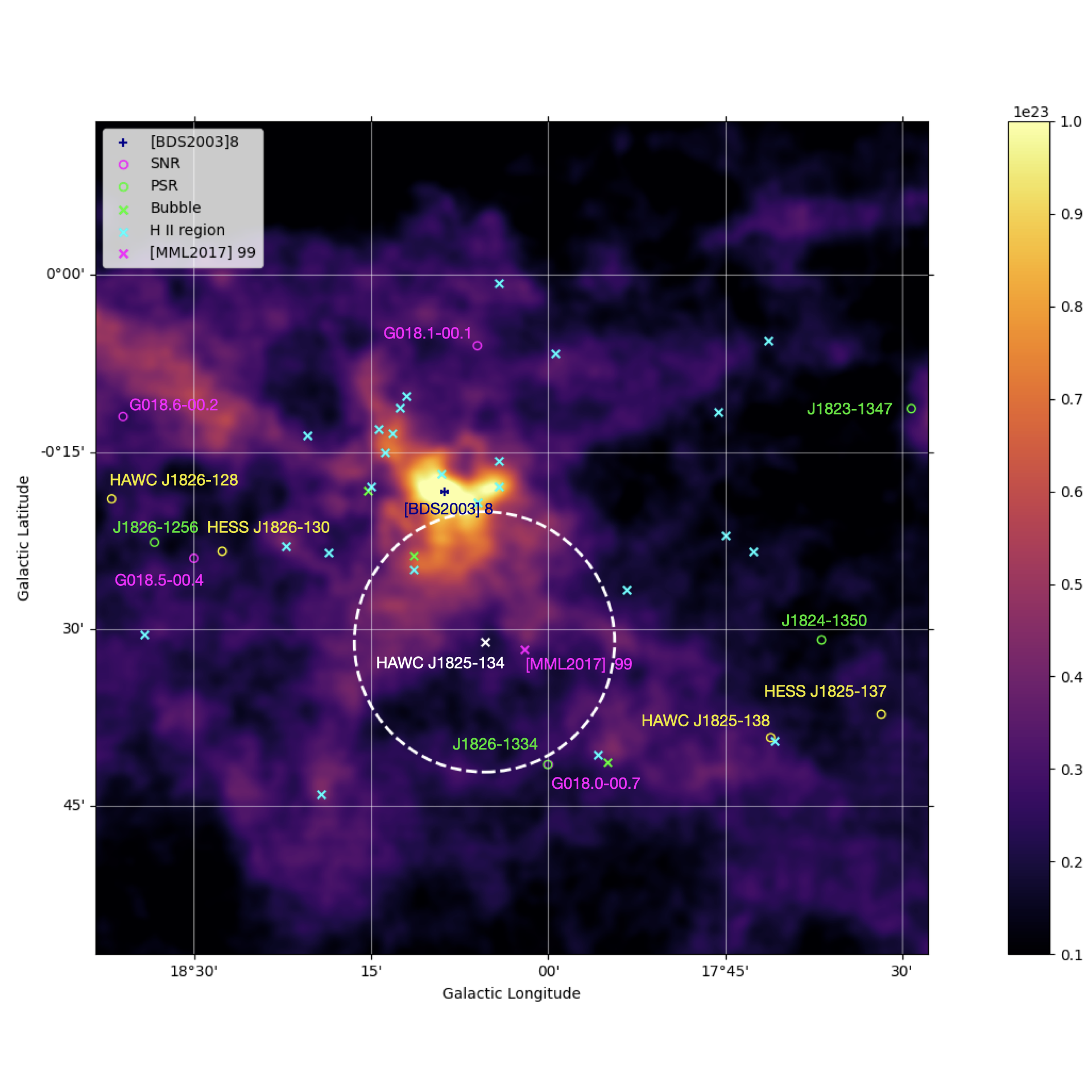}%%
}
\caption {FUGIN $^{12}$CO molecular column density in units of cm$^{-2}$ obtained by integrating the line intensity over a range in velocity between 40 to 60 km/s, corresponding to a distance for the molecular gas of 3-4 kpc. The best fitted location of the new HAWC source and the upper limit to its extension, obtained for all events above 1 TeV, are indicated with a white x and white dashed circle, respectively. The yellow circles represent the positions of the HESS~J1825-137, HESS~J1826-130, HAWC~J185-137 and HAWC~J1826-128. The giant molecular cloud, GMC 99, from \citep{Miville} catalogue is the magenta x. The cloud angular extension is 0.36$^\circ$ in radius and its total mass amounts to 4.5 $\times$ 10$^{5}$ solar masses. The new HAWC source is compatible with the location and extension of this cloud.  The location of the young star cluster IRC [BDS2003] 8 is marked with a blue + \citep{Bica:2003kf,2016A&A...585A.101K}. Green circles represent the locations of known pulsars from the ATFN catalogue \citep{Manchester:2004bp} and magenta circles the locations of supernova remnants \citep{Ferrand:2012jh}. Green and cyan x indicate bubbles from \citep{Churchwell:2006jf} and HII regions from the WISE Catalogue by \citep{Anderson:2014}, respectively.}
\label{Fig3}
\end{center}
\end{figure}

In Figure~\ref{Fig3}, the distribution of ambient molecular gas in the region of HAWC J1825-134 is shown by integrating the intensity of the $^{12}$CO(1-0) line measured by the FUGIN survey \citep{FUGIN} over a range of velocities between 40 and 60 km/s corresponding to the peak in the gas spectrum from the HAWC J1825-134 region (see Figure~\ref{COSpectrum} in Section~\ref{sec:gasdensity}). Using the estimation of radial velocity as a function of direction (due to Galactic rotation) the range between 40 and 60 km/s radial velocity at Earth corresponds to a distance between 3 and 4 kpc \citep{Blitz}.
%The $^{13}$CO line intensity traces the colder phase of the molecular gas aggregated in the cloud clumps. 
The hundred TeV emission is associated with a very bright peak in the distribution of the molecular gas, which is consistent with a hadronic scenario for the origin of the emission at the highest energies. Furthermore, the best fitted position of HAWC J1825-134 is compatible with the position of the giant molecular cloud 99 of the \citep{Miville} catalogue (see Figure~\ref{Fig3}). The cloud with an angular extension of 0.36$^{\circ}$ has a radial velocity of 48 km/s corresponding to 3.9 kpc near distance and a mass of 4.5 $\times$ 10$^5$ solar masses. 

In Figure~\ref{Fig3} the best fitted position of the new HAWC source is shown in white cross. Known pulsars, SNRs, stellar clusters, and molecular clouds are marked on the plot. The SNR~G018.1-00.1 or the SN progenitors of the pulsars, PSR J1826-1334 or PSR J1826-1256, are not likely the sources of PeV protons because of their ages of 5-9 kyr, 21 kyr and 14 kyr, respectively \citep{Leahy:2013,Ferrand:2012jh}. PeV protons 
accelerated in these SNRs would have since escaped the whole region, unless the particle diffusion coefficient were locally suppressed by a factor of 1000 for G018.1-00.1 (assuming an age of 5 kyr \citep{Leahy:2013}) or more for the others. Additionally the distance of G018.1-00.1 is 6.4 $\pm$ 0.4 kpc \citep{Leahy:2013} and the SNR G018.26-00.2 (SNR distance of 4.6 $\pm$ 0.2 kpc) is located beyond 0.5$^o$ from the centre of the HAWC emission. 

As suggested by \citep{Amato:2006ts} TeV protons could be accelerated in pulsars. PeV protons producing the radiation from HAWC J1825-134 could be accelerated in one of the nearby powerful pulsars, PSR~J1826-1334 or PSR~J1826-1256. These pulsars have a total energy budget amounting to roughly 10$^{48}$ ergs (the spin down luminosity of both pulsars is roughly 3 $\times$ 10$^{36}$ erg/s and their age 15-20~kyr). Roughly 10$\%$ of the total energy could account for the total energy budget in CRs from HAWC J1825-134 above 1 TeV. The spin down luminosity might have been higher in previous phases of the pulsars. This would increase the total budget available. On the other hand assuming a spherically symmetric diffusion of the accelerated hadrons from the pulsars, one expects that the trajectory of only a small fraction of these particles would lead them to collide with the gas in the cloud, making the required energy output in CRs from the local pulsars higher than the total spin down luminosity. 

Located near the HAWC new point source, the 2MASS young star cluster, [BDS2003] 8, indicated in Figure~\ref{Fig3} with a blue cross, is about 1 million years old and at a distance of about 4 kpc from the Earth. The cluster is thus at the same distance from us as GMC 99 \citep{Bica:2003kf,2016A&A...585A.101K}. The typical total kinetic energy 
 of OB winds in young star clusters is about 5 $\times$ 10$^{38}$ erg/s, which over the cluster lifetime of 1 million years produces a total of 1.5 $\times$ 10$^{52}$ ergs. A very small fraction of this energy needs to be channeled into proton acceleration to explain the emission measured from HAWC J1825-134. Both from energetic considerations and from its position, [BDS2003] 8 is the most likely source of the PeV cosmic rays producing HAWC emission. The presence of several turbulent cloud clumps, bubbles (green crosses in Figure~\ref{Fig3}) and HII regions (cyan crosses in Figure~\ref{Fig3}) within HAWC J1825-134 is likely to favor the confinement of PeV cosmic rays, which would otherwise quickly escape the region.  The discovery of HAWC~J1825-134 supports the hypothesis that young star clusters are important contributors to the Galactic CR spectrum up to the knee as argued recently by \citep{Aharonian:2018oau,Hona:2019ysf}.

Finally high energy neutrinos are expected to be produced in the same proton-proton collisions which produce hundred TeV $\gamma$-rays. Indeed, the 1040 TeV event 14 in the public Ice Cube high energy event list \citep{Aartsen:2014gkd} is compatible with originating from the HAWC~J1825-134 region. However, due to the angular resolution of this Ice Cube shower event (10-15 deg), it is not possible to conclusively associate it to the HAWC source. On the other hand, first estimations of the detectability of this source for the upcoming northern neutrino telescope, KM3NeT, are promising \citep{Niro:2019mzw}.

%%%%%%%%%%%%%%%%%%%%%%%%%%%
%%%%%%%%%%%%%%%%%%%%%%%%%%%

\section{Conclusions} 
\label{sec:conclusions}
In summary we report the discovery of the $\gamma$-ray source, HAWC J1825-134, whose spectrum extends well beyond 200 TeV without evidence of a cutoff or break in the spectrum. The source is found to be coincident with a region where the density is as high as 700 protons/cm$^3$ and with the giant molecular cloud 99 of \citep{Miville} catalogue. Although from HAWC data we can not exclude that the new source is coincident with the pulsar PSR J1826-1334 location, we found that this scenario is disfavored. Therefore PeV cosmic ray protons colliding with the ambient gas likely produce the observed radiation. We argue that the possible source of these multi PeV cosmic rays is the young star cluster, [BDS2003] 8. The high CR energy density needed to explain the emission can result from effective CR confinement within the turbulent region of HAWC~J1825-134.
While a definitive association to this young stellar cluster is still to be clarified, HAWC J1825-134 provides the $\gamma$-ray signature of an extreme accelerator along the Galactic Plane and will be an important target for the upcoming neutrino experiments, KM3NeT~\citep{Adrian-Martinez:2016fdl} and IceCube-Gen2~\citep{Aartsen:2014gkd}, and new $\gamma$-ray observatories like LHAASO Observatory~\citep{Bai:2019khm}, the SWGO Observatory~\citep{Abreu:2019ahw}, and the Cherenkov Telescope Array~\citep{CTAConsortium:2018tzg}.

\acknowledgments

We acknowledge the support from: the US National Science Foundation (NSF); the US Department of Energy Office of High-Energy Physics; the Laboratory Directed Research and Development (LDRD) program of Los Alamos National Laboratory; Consejo Nacional de Ciencia y Tecnolog\'ia (CONACyT), M\'exico, grants 271051, 232656, 260378, 179588, 254964, 258865, 243290, 132197, A1-S-46288, A1-S-22784, c\'atedras 873, 1563, 341, 323, Red HAWC, M\'exico; DGAPA-UNAM grants IG101320, IN111315, IN111716-3, IN111419, IA102019, IN112218; VIEP-BUAP; PIFI 2012, 2013, PROFOCIE 2014, 2015; the University of Wisconsin Alumni Research Foundation; the Institute of Geophysics, Planetary Physics, and Signatures at Los Alamos National Laboratory; Polish Science Centre grant, DEC-2017/27/B/ST9/02272; Coordinaci\'on de la Investigaci\'on Cient\'ifica de la Universidad Michoacana; Royal Society - Newton Advanced Fellowship 180385; Generalitat Valenciana, grant CIDEGENT/2018/034; Chulalongkorn University’s CUniverse (CUAASC) grant. Thanks to Scott Delay, Luciano D\'iaz and Eduardo Murrieta for technical support.

This publication makes use of molecular line data from the Boston University-FCRAO Galactic Ring Survey (GRS). The GRS is a joint project of Boston University and Five College Radio Astronomy Observatory, funded by the National Science Foundation under grants AST-9800334, AST-0098562, AST-0100793, AST-0228993, and AST-0507657.

This publication makes use of data from FUGIN, FOREST Unbiased Galactic plane Imaging survey with the Nobeyama 45-m telescope, a legacy project in the Nobeyama 45-m radio telescope.





\newpage
\appendix
\section{Modeling studies}
\label{sec:modeling}

The model of the analysis includes two extended two-dimensional Gaussian-shaped sources (HAWC~J1825-138 and HAWC~J1826-128), one point source (HAWC~J1825-134), and an extended GDE component. Other morphologies have been tested, in particular for HAWC~J1825-138 a disk-shaped extended and an asymmetric Gaussian. However, none of these models yields a better agreement with data (see Table~\ref{tab: table2}). We also tried an extended source hypothesis for HAWC~J1825-134, however the 3ML fit gives large uncertainties for the fitted parameters so instead an upper limit to the extension was calculated, 0.18$^{\circ}$ at 95\% CL.

We also tried different spectral assumptions. For a simple power law the 3ML fit does not converge, which is expected based on the curved trend shown by the flux points (see Figure~\ref{fig: spectra} in the main text). A log-parabola (see for instance Eq. 2 in~\citep{Abeysekara:2019gov} as a reference for a log-parabola spectrum) was tested both for HAWC~J1825-138 and HAWC~J1826-128 (see Table~\ref{tab: table2}). Finally a simple power law with an energy cut-off for HAWC~J1825-134 was tested but again the 3ML fit does not converge.
We use the Bayesian Information Criterion (BIC)~\citep{Schwarz:1978tpv} to compare the different models tested. The values are summarized in Table~\ref{tab: table2}.
Here, $\Delta$BIC = BIC$_{\rm{test}}$ - BIC$_{\rm{ref}}$ is the difference in BIC between the reference and tested model. The best model is the one minimizing the BIC value.

%%%%%%%%%  TABLE  %%%%%%%%%%%%
\begin{table}[htpb]
%\scriptsize
%\footnotesize
%\small
%\renewcommand{\arraystretch}{2}
  \begin{center}
    \begin{tabular} { |c|c|} 
      \hline
      Morphological model & $\Delta$BIC \\
      \hline
      HAWC J1825-138 Gaussian & 0 \\
      \hline
      HAWC J1825-138 disk & 37 \\
      \hline
      HAWC J1825-138 asymmetric Gaussian & 23 \\ 
      \hline
      \hline
      Spectral model & $\Delta$BIC \\
      \hline
      HAWC J1825-138 Ecut + HAWC J1826-128 Ecut & 0 \\
      \hline
      HAWC J1825-138 logP + HAWC J1826-128 Ecut & 4 \\
      \hline
      HAWC J1825-138 Ecut + HAWC J1826-128 logP & 1.4 \\
      \hline
    \end{tabular}
  \end{center}
  \caption{Summary of the models tested with the corresponding $\Delta$BIC.}
  \label{tab: table2}
\end{table}
%%%%%%%%%%%%%%%%%%%%%%%%%%%%%%

In order to check the possibility that HAWC~J1825-134 comes from statistical fluctuations from either HAWC~J1825-138 and/or HAWC~J1826-128 we performed a Monte Carlo study where we simulated more than 2600 synthetic maps using the best fitted two-source model and Poisson fluctuations. Then we fitted with 3ML the simulated data sets with both the two-source model and the three-source model. None of these simulations reached the observed $\Delta$TS value, computed according to Eq.~\ref{dts}. To asses the statistical significance of the result, instead of using additional simulations, which is computationally expensive, we estimated the p-value of the observed $\Delta$TS extrapolating the simulated $\Delta$TS distribution. The simulation result shows that the three-source model is preferred at the 5.2\,$\sigma$ level, showing that the source detection is not likely produced by a source contamination or statistical fluctuations.

Finally, we also found indications of energy dependent morphology for HAWC~J1825-138, showing that the higher the energy the smaller the size. However, with the current statistical and systematic uncertainties this cannot be claimed at the moment. We expect that with a larger HAWC dataset it can be confirmed.

\section{Uncertainty in the source location and leptonic mechanism}
\label{sec:location}

Here we show why the conclusions on the severe electron cooling do not thus change when considering the uncertainty in the location of HAWC J1825-134. Accounting for the statistical and systematic uncertainties in the fitted location of the source, HAWC J1825-134, mentioned in the previous section, we checked that the column density within half a degree around the position of HAWC J1825-134 varies between 2.5 and 5 $\times$ 10$^{22}$ cm$^{-2}$, so that the gas density is n $\geq$ 600 cm$^{-3}$ at 4 kpc, the magnetic field in the region, B $\geq$ 24 $\mu$Gauss, and the electron cooling time, t$_{cool} \leq$ 30 yr.  If the electrons could travel at light speed they would cover at most the distance of 10 pc. In reality, the distance travelled should be orders of magnitude smaller if the severe suppression of the diffusion coefficient in dense gas regions is taken into account. The conclusions on the severe electron cooling do not thus change when considering the uncertainty in the location of HAWC J1825-134.

Finally let us assume that HAWC J1825-134 is coincident with the position of the pulsar, PSR J1826-1334. In this case one could think that the emission from HAWC J1825-134 is produced through IC scattering off CMB photons by high energy electrons accelerated in loco by the pulsar, if these are not cooled down through synchrotron emission. In order to constrain the magnetic field and thus the synchrotron losses in the region of HAWC J1825-134 we use Suzaku measurements of the X-ray pulsar associated to PSR J1826-1334. The magnetic field B inside the 0.15 deg extended X-ray nebula is 7 $\mu$G~\citep{Uchiyama:2008bd}.  Assuming that the same electron population producing the X-ray nebula emits also through IC above several tens of TeV, the predicted IC flux would be one order of magnitude lower than the flux measured by HAWC, following L$_{\gamma}/L_X \sim$  0.1(B/10 $\mu$G) \citep{Aharonian:2004yt}. We conclude that HAWC J1825-134 is unlikely to be produced by the pulsar, PSR J1826-1334.

%%%%%%%%%%%%%%%%%%%%%%%%%%%
\begin{figure}[htpb]
\begin{center}
\resizebox{1.0\textwidth}{!}{%
\includegraphics[width=0.5\textwidth]{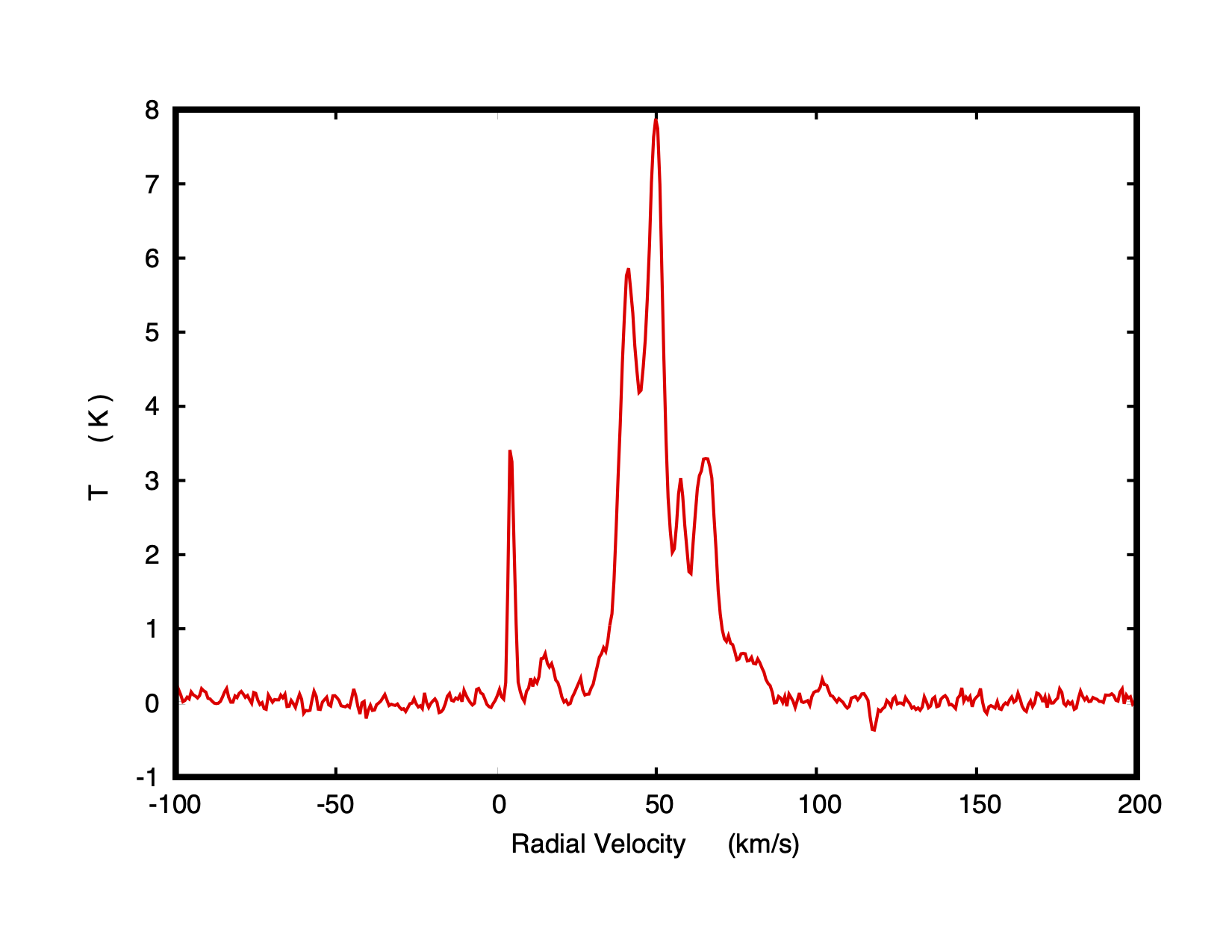}%
}
\caption{FUGIN $^{12}$CO(1-0) radial velocity spectrum from the HAWC J1825-134 region.\citep{FUGIN}.}
    
\label{COSpectrum}
\end{center}
\end{figure}   
%%%%%%%%%%%%%%%%%%%%%%%%%%%%

\section{Gas Distribution}
\label{sec:gasdensity}

In order to estimate the total gas target for the PeV protons within HAWC J1825-134, 3D (l-b-v) cubes from the FUGIN survey of $^{12}$CO(1-0) emission line \citep{HI4PI:2016,FUGIN,Voisin:2016mdj} and from the HI4PI survey of atomic hydrogen \citep{HI4PI:2016} are integrated over the same range in velocity 40-60 km/s. This velocity range corresponds to a peak in the FUGIN $^{12}$CO(1-0) radial velocity spectrum as observed from Earth of HAWC J1825-134 region in Figure~\ref{COSpectrum}. 

A conversion factor X=1.8 $\times {10}^{20}$ cm$^{-2}$ K$^{-1}$ km$^{-1}$ has been used to translate the $^{12}$CO(1-0) column density in molecular hydrogen column density \citep{Dame:2000sp}. The uncertainty on the conversion factor, which is of 30 $\%$, is also the source of uncertainty on the gas mass. The average gas column density from the region of HAWC J1825-134 amounts to about  N = 3 $\times$ 10$^{22}$ cm$^{-2}$, of which about 10 $\%$ is atomic hydrogen.

The ambient gas density is 700 cm$^{-3}$, calculated under the assumption that the volume of HAWC J1825-134 is a sphere of radius r = 13 pc, corresponding to an extension of 0.18$^{\circ}$ (upper limit) at a distance of 4 kpc. The total gas mass within HAWC J1825-134 amounts to about 1.6 $\times$ 10$^5$ solar masses, of which 1-2 10$^4$ solar masses is contained in cold clumps. An estimate of the column density in cold clumps within the region of the new HAWC source can be obtained by making use of the intensity of the $^{13}$CO(1-0) line measured by the Galactic Ring Survey (GRS) \citep{Jackson:2006ye,Simon:2001zz}. This line intensity traces in fact the colder phase of the molecular gas aggregated in very dense gas regions and cloud clumps. The integration of the line over the range of velocities between 40 and 60 km/s yields an average column density of 2 $\times$ 10$^{21}$ cm$^{-2}$, corresponding to a total mass in cold clumps of 1.5 $\times$ 10$^4$ solar masses.

As we mentioned r=13 pc is an upper limit for the size of HAWC J1825-134 and the gas column density, N = $\int n \, dl$, in a region of half a degree around the fitted position of HAWC J1825-134 varies between 2.5 and 5 $\times$ 10$^{22}$ cm$^{-2}$. Since the gas density is roughly n $ \propto N/r$, a smaller size of the HAWC source, r', would roughly increase the gas density by a factor 13pc/r'. A higher gas density would further support the main conclusion of the paper, namely the hadronic origin of the HAWC emission.

\section{Observation of the HAWC J1825-134 region by other $\gamma$-ray observatories}
\label{sec:observatories}
The region of eHWC J1825-134 was observed in TeV photons by H.E.S.S. and VERITAS, two air Cherenkov $\gamma$-ray observatories. The H.E.S.S. collaboration has detected two $\gamma$-ray sources, HESS J1825-137 ~\citep{Abdalla:2018qgt} and HESS J1826-130 ~\citep{Abdalla:2020dmz} in the region, whose location, morphology and energy spectra are consistent within uncertainties with HAWC J1825-138 and HAWC J1826-128. The source HAWC J1825-134 is however not detected by H.E.S.S., possibly because of its high-energy nature. Moreover, even if it is considered as point source, a small spatial extension, up to 0.18$^{\circ}$, is not excluded by this analysis which would make its detection more challenging for air Cherenkov observatories. 
%Interestingly, H.E.S.S. data may present some hints of hard emission in the location reported for HAWC J1825-134. In particular, in~\citep{Abdalla:2018qgt} high-energy emission (Fig.~3, E>10~TeV) is observed in the location of HAWC J1825-134. Moreover, this location coincides with the hardest spectral emission as displayed in Fig.~8 of the same publication.
Finally, the region has also been observed by the VERITAS collaboration~\citep{Abeysekara:2020enu}, but with smaller exposure time, and therefore, less sensitivity than the H.E.S.S. observation.



\end{document}